\documentclass{pasj00}

\begin{document}
\SetRunningHead{Miura et al.}{Secular increase of the Astronomical Unit}
\Received{2009/06/23}
\Accepted{2009/08/01}

\title{
Secular increase of the Astronomical Unit: 
   \\ a possible explanation in terms of \\ 
   the total angular momentum conservation law
}

\author{
Takaho \textsc{Miura}\altaffilmark{1}
Hideyoshi \textsc{Arakida}\altaffilmark{2}
Masumi \textsc{Kasai}\altaffilmark{1}
and
Shuichi \textsc{Kuramata}\altaffilmark{1}
}
\altaffiltext{1}{Graduate School of Science and Technology, 
   Hirosaki University, Hirosaki Aomori 036-8561, Japan}
\altaffiltext{2}{ School of Education, Waseda University, Shinjuku, 
	 Tokyo 169-8050, Japan}


%


\maketitle

\begin{abstract}
We give an idea and the order-of-magnitude estimations to explain the
recently reported secular increase of the Astronomical Unit (AU) by
Krasinsky and Brumberg (2004). The idea
proposed is analogous
   to the tidal acceleration in the Earth-Moon system, which is based
   on the conservation of the total angular momentum and 
   we apply this scenario to the Sun-planets system.
    Assuming the existence of some tidal interactions that 
    transfer the rotational angular momentum of the Sun
   and using reported value of the positive secular trend 
   in the astronomical unit,
  $\frac{d}{dt}\mbox{AU} = 15 \pm 4\, \mbox{(m/cy)}$, the suggested
  change in the period of rotation of the Sun is about $21\,
  \mbox{ms/cy}$ in the case that the orbits of the eight planets have
  the same ``expansion rate.''
  This value is sufficiently small, and at present it seems 
  there are no observational data which exclude this possibility.
  Effects of the change in the Sun's moment of inertia
  is also investigated.
  It is pointed out that the change in the moment of inertia 
  due to the radiative mass loss by 
  the Sun may be responsible for the secular increase of AU, if the
  orbital ``expansion'' is happening only in the inner planets system.
Although the existence of some tidal interactions is assumed between
the Sun and planets, concrete mechanisms of the angular momentum
transfer are not discussed in this paper, which remain to be done as
future investigations.

\end{abstract}
\section{Introduction}
The Astronomical Unit (hereafter we abbreviate AU) is one
of the most essential scale in astronomy which characterizes
the scale of the solar system and the standard of cosmological 
distance ladder. 
AU is also the fundamental astronomical constant that 
associates two length unit; one $\mbox{(m)}$ in International 
System (SI) of Units and one $\mbox{(AU)}$ in Astronomical 
System of Units. 

In the field of fundamental astronomy e.g., the planetary
ephemerides, it is one of the most important subjects to 
evaluate AU from the observational data. 
The appearance of planetary radar and spacecraft ranging
techniques has led to great improvement of determination of AU.
Modern observations of the major planets, including the
planetary exploration spacecrafts such as Martian landers and 
orbiters, make it possible to control the 
value of AU within the accuracy of one meter or even better. 
Actually present best-fit value of AU is obtained as
(\cite{pitjeva2005}),
\begin{equation}\label{eq:au1}
1~\mbox{(AU)} = {\rm AU}~ \mbox{(m)} =
1.495978706960 \times 10^{11} \pm 0.1~ \mbox{(m)}.
\end{equation}

However, recently Krasinsky and Brumberg reported the positive 
secular trend in AU as 
$\frac{d}{dt}\mbox{AU} = 15 \pm 4\, \mbox{(m/cy)}$
\footnote{In this paper, $\mbox{cy}$ means the century according 
to Krasinsky and Brumberg (\cite{kb2004}).} 
from the analysis of radar ranging of inner planets and
Martian landers and orbiters (\cite{kb2004,standish2005}).
As mentioned above, the value of AU is presently obtained 
within the error of one meter or even better, so that
the reported $d{\rm AU}/dt = 15 \pm 4\, \mbox{(m/cy)}$ seems 
to be in excess of current determination error of AU.
This secular increase of AU 
was discovered by using following formula
(\cite{krasinsky2007}),
\begin{equation} \label{eq:au2}
 t_{\rm theo} = \frac{d_{\rm theo}}{c}
  \left[
   \mbox{AU} + \frac{d\mbox{AU}}{dt}(t - t_0)
  \right],
\end{equation}
in which $t_{\rm theo}$ is the theoretical value of round-trip 
time of radar signal in the SI second, $d_{\rm theo}$ is 
interplanetary distance obtained from ephemerides 
in the unit of $\mbox{(AU)}$, 
$c$ is the speed of light in vacuum, $\mbox{AU}$ and 
$d\mbox{AU}/dt$ are, respectively, the astronomical unit as 
Eq. (\ref{eq:au1}) 
and its time variation, and $t_0$ is the initial epoch of 
ephemerides. $t_{\rm theo}$ is compared with
the observed lapse time of signal $t_{\rm obs}$ and then 
$\mbox{AU}, d\mbox{AU}/dt$ are fit by the least square method. 

%
It should be emphasized that, at this time, the existence of secular
increase of AU is not confirmed robustly by other ephemerides
groups, 
and there is still room for the independent analysis of observational data.
However, several attempts have already been made to explain this
possible secular
increase of AU, including e.g., the effects of the cosmic expansion
(\cite{kb2004,mashhoon2007,arakida2009}), mass loss of 
the Sun (\cite{kb2004}, \cite{noerdlinger2008}), 
the time variation of gravitational constant $G$ (\cite{kb2004}), 
the influence of dark matter (\cite{arakida2008}) and so on.
But unfortunately so far none of them seems to be successful.

In this paper, we will take another viewpoint and give 
an explanation for the secular increase of AU, based on the 
standard conservation law of the total angular momentum. 
The mechanism is closely analogous to the case of the tidal
acceleration in the Earth-Moon system 
which induces the transfer of the rotational angular momentum of 
the Earth into the orbital angular momentum of the Moon due to 
tidal friction. 
We will apply similar scenario to the 
Sun-planets system.  
\section{Tidal acceleration in the Earth-Moon system: a brief summary} 

In this section, we briefly summarize the tidal acceleration in the
Earth-Moon system. See e.g., Chapter 4 of 
\citet{murray-dermott} for details.

The conservation law of the total angular momentum in the Earth-Moon
system is 
\begin{equation}\label{eq:cons}
  \frac{d}{dt}\left(\ell_{\mbox{\scriptsize E}} + L_{\mbox{\scriptsize
        M}} \right) = 0, 
\end{equation}
where $\ell_{\mbox{\scriptsize E}}$ is the rotational angular momentum of the
Earth, 
$L_{\mbox{\scriptsize M}}$ is the orbital angular momentum of the Moon, and for
simplicity we have neglected the rotational angular momentum of the
Moon, which is about $10^{-5}$ times smaller than
$L_{\mbox{\scriptsize M}}$. 
The rotational angular momentum is written in terms of the moment of
inertia $I_{\mbox{\scriptsize E}}$ and the period of rotation
$T_{\mbox{\scriptsize E}}$ as 
\begin{equation}
  \ell_{\mbox{\scriptsize E}} = I_{\mbox{\scriptsize E}}\,
  \frac{2\pi}{T_{\mbox{\scriptsize E}}}.  
\end{equation}
The moment of inertia can be written as
\begin{equation}
  I_{\mbox{\scriptsize E}} = \gamma M R^2, 
\end{equation}
where $M$ is the mass and $R$ is the radius of the Earth, 
and $\gamma$
is the moment of inertia factor. In the case of the perfect sphere with
uniform density inside, $\gamma$ is $2/5$.  
For a centrally-condensed body, $\gamma$ is in general less than
$2/5$. 
Often quoted value for the Earth is 
$\gamma = 0.3308$. 
The orbital angular momentum of the Moon is written as 
\begin{equation}
  L_{\mbox{\scriptsize M}} =  m \sqrt{GM r (1-e^2)}, 
\end{equation}
where $m$ is the mass of the Moon, 
$r$ and $e$ is the semi-major axis
and the eccentricity of the Moon's orbit, respectively, 
and $G$ is the
gravitational constant. 
For the sake of simplicity, hereafter we assume $e$ is constant. 

Then, if we assume $I_{\mbox{\scriptsize E}}$, $m$, and $M$ are
constant, the conservation of the total angular momentum
Eq.~(\ref{eq:cons}) reads
\begin{equation}\label{eq:5}
  \frac{\dot{T}_{\mbox{\scriptsize E}}}{T_{\mbox{\scriptsize E}}} 
  = \frac{1}{2} \frac{L_{\mbox{\scriptsize
        M}}}{\ell_{\mbox{\scriptsize E}}} \frac{\dot r}{r}   
\end{equation}

The motion of the Moon can be traced with an accuracy of a few
centimeters by the lunar laser ranging experiments. 
The measurements yield the numerical
value $\dot r = +3.82 \pm 0.07 \,\mbox{(m/cy)}$ (\cite{dickey1994})  
then 
\begin{equation}\label{eq:6}
\frac{\dot r}{r} \simeq 1.0\times 10^{-8}\,(\mbox{cy}^{-1}).   
\end{equation}
Using Eq.~(\ref{eq:6}), the numerical value of
Eq.~(\ref{eq:5}) is
\begin{equation}
  \frac{\dot{T}_{\mbox{\scriptsize E}}}{T_{\mbox{\scriptsize E}}}
 \simeq 
8.1 \times \gamma^{-1} \times 10^{-9} \,(\mbox{cy}^{-1}), 
\end{equation}
or equivalently, $\dot{T}_{\mbox{\scriptsize E}} \simeq
2.1\,(\mbox{ms/cy})$, where we have used 
$\gamma = 0.3308$. 

On the other hand, historical records over 2700 years can be used to
compute the change in the length of the day, and the following average
value is found:
\begin{equation}
  \dot T_{\mbox{\scriptsize obs}} = +1.70 \pm 0.05\,(\mbox{ms/cy}). 
\end{equation}
The gradual slowing of the Earth's rotation is due to the tidal force
between the orbiting Moon and the Earth, or the tidal friction. 
It is explained that the discrepancy between the estimated value 
$\dot{T}_{\mbox{\scriptsize E}}$ and the observed one 
$\dot T_{\mbox{\scriptsize obs}}$ is largely due to the change in the
moment of inertia of the Earth caused by the melting of ice at the
poles. 
%
\section{Application to the Sun-planets system}
In this section, we apply the same argument in the previous section to
the Sun-planets system.  
All we need is the conservation of the total angular momentum in the
solar system. 
We denote the mass and the orbital elements of each planet by
subscript $i$. 
The length AU,
denoted by $a$, is used to normalize $r_i$, then for the Earth's radius
$r_3 \equiv r_{\mbox{\scriptsize E}} = a$, and for the moment it is assumed that
the orbits of the all planets have the same ``expansion rate,'' i.e., 
\begin{equation}\label{eq:exprate}
  \frac{\dot r_i}{r_i} = \frac{\dot a}{a}
\end{equation}
for each $i$. 

In the case that the Sun's moment of inertia does not change, from the
analogy to Eq.~(\ref{eq:5}), we obtain the change in the period of
rotation of the Sun $T_{\odot}$ as
\begin{equation}\label{eq:10}
  \frac{\dot T_{\odot}}{T_{\odot}} =
  \frac{1}{2} \frac{L}{\ell_{\odot}} \frac{\dot a}{a}, 
\end{equation}
where 
$L$ is the sum of the orbital angular momentums of all planets
\begin{equation}
  L = \sum_i m_i \sqrt{GM_{\odot} r_i
      (1-e_i^2)}, 
\end{equation}
$\ell_{\odot}$ is the rotational angular momentum of
the Sun 
\begin{equation}
  \ell_{\odot} = I_{\odot} \,
  \frac{2\pi}{T_{\odot}}=
 \gamma_{\odot}
  M_{\odot} R^2_{\odot} \, \frac{2\pi}{T_{\odot}}, 
\end{equation}
and $M_{\odot}$, $R_{\odot}$, and $\gamma_{\odot}$ are the
mass, the radius, and 
the moment of inertia factor of the Sun, 
respectively. 

Using the value $\dot a \simeq 15 (\mbox{m/cy})$ reported by Krasinsky
and Brumberg, $\dot a/a \simeq 1.0\times 10^{-10}$ and
the right-hand-side of Eq.~(\ref{eq:10}) is evaluated as
\begin{equation}\label{eq:13}
  \frac{\dot T_{\odot}}{T_{\odot}}
  \simeq
  5.7 \times \gamma^{-1}_{\odot}\times 10^{-10}
\,(\mbox{cy}^{-1}).
\end{equation}
If we use the value 
$\gamma_{\odot}=0.0059$ and the
rotational period of the Sun as $T_{\odot} =
25.38\,(\mbox{days})$, the estimated value of the change in 
$T_{\odot}$ is
\begin{equation}
  \dot T_{\odot} \simeq 
21
\,(\mbox{ms/cy}). 
\end{equation}
The estimated value is sufficiently small and seems to be well within the
observational limit. 
\section{Effects of the change in the moment of inertia}
In the previous section, we have considered only the case that the
moment of inertia $I_{\odot}$ is constant, namely, 
\begin{equation}
  \frac{d}{dt} \ell_{\odot} = 
  \frac{d}{dt}
  \left(I_{\odot}\,\frac{2\pi}{T_{\odot}}\right)
  = - \frac{\dot T_{\odot}}{T_{\odot}} \ell_{\odot}\, . 
\end{equation}

If we generalize our result to the case that $I_{\odot}$ and
$M_{\odot}$ are not
constant, Eq.~(\ref{eq:10}) changes to 
\begin{equation}\label{eq:16} 
  - \frac{\dot \gamma_{\odot}}{\gamma_{\odot}}
  -\frac{\dot M_{\odot}}{M_{\odot}}
  -2\frac{\dot R_{\odot}}{R_{\odot}}
  + \frac{\dot T_{\odot}}{T_{\odot}} =
    \frac{1}{2} \frac{L}{\ell_{\odot}}
\left(\frac{\dot M_{\odot}}{M_{\odot}} +  \frac{\dot a}{a}\right)\,.
\end{equation}
As a first approximation, we assume that the radiative mass loss
occurs isotropically along radial direction and does not carry the
angular momentum.

The first term in the left-hand-side of Eq.~(\ref{eq:16}) represents
the effect of change in the internal density distribution of the Sun,
and so far we do not have enough information on it in detail. 

The second term in the left-hand-side of Eq.~(\ref{eq:16}) represents
the effect of mass loss, which can be evaluated in the following way.
The Sun has luminosity at least $3.939\times 10^{26}\,\mbox{W}$, or
$4.382\times 10^{9}\,\mbox{kg/s}$, including electromagnetic radiation
and contribution from neutrinos (\cite{noerdlinger2008}). 
The particle mass loss rate
by the solar wind is about $1.374\times 10^9 \,\mbox{kg/s}$, according
to \citet{noerdlinger2008}. The total solar mass 
loss rate is then 
\begin{equation}\label{eq:9.1}
  -\frac{\dot M_{\odot}}{M_{\odot}} = 
   9.1 \times 10^{-12} \,(\mbox{cy}^{-1})\, , 
\end{equation}
which is less than a thousandth
 of the required value to explain the
secular increase of AU (see Eq.~(\ref{eq:13})).  
(As pointed out by Noerdlinger (2008), Krasinsky and Brumberg (2004) unaccountably ignored the
radiative mass loss $L_\odot = 3.86 \times 10^{26}\, \mbox{W}$ which
is the major contribution to 
$\dot{M_{\odot}}/{M_{\odot}}$.)
Therefore, we can
conclude that the solar mass loss term in the left-hand-side of
Eq.~(\ref{eq:16}) does not make a significant contribution to the
secular increase of AU, if the orbits of the eight planets have the same
``expansion rate.''

Note that the term which is proportional to $\dot M_{\odot}/M_{\odot}$
also appears in the right-hand-side of Eq.~(\ref{eq:16}). This term
may be called as the Noerdlinger effect (\cite{noerdlinger2008}). 
Noerdlinger already investigated this 
effect of solar mass loss, and concluded that the effect
can only account for less than a tenth of the reported value by
Krasinsky and Brumberg. 

The third term in the left-hand-side of Eq.~(\ref{eq:16}) is the
contribution from the change in the solar radius. Although the very
short-time and small-scale variability in the solar radius may be
actually observed in the context of helioseismology, we have no
detailed information on the secular change in $R_{\odot}$ so far. 
\section{Case of the Sun-inner planets system}
In the previous sections, we have assumed that the orbits of all the 
planets have the same ``expansion rate'', i.e.,
Eq.~(\ref{eq:exprate}).  However, the recent positional observations
of the planets with high accuracy are mostly done within the inner planets
region.  
Actually Krasinsky and Brumberg obtained $d{\rm AU}/dt$ by using  
these inner planets data, while the orbits of outer planets are
given by the Russian ephemeris EPM (Ephemerides of Planets and the
Moon).
Therefore, as a tentative approach, we 
consider the case that the ``expansion rates'' of
the planetary orbits are not homogeneous but inhomogeneous in the
sense that only the orbits of the inner planets expand.
 
In this section, we consider the case that the
``expansion'' of the planetary orbit occurs only for the inner
planets. 

In this case,  the sum of the orbital angular momentum of all
planets $L$ is replaced by the sum of the inner planets
$L_{\mbox{\scriptsize in}}$:  
\begin{equation}
  L_{\mbox{\scriptsize in}} \equiv \sum_{i=1}^4  m_i 
   \sqrt{GM_{\odot} r_i(1 - e^2_i)}. 
\end{equation}
Then Eq.~(\ref{eq:16}) is now 
\begin{equation}\label{eq:22}
    - \frac{\dot \gamma_{\odot}}{\gamma_{\odot}}
  -\frac{\dot M_{\odot}}{M_{\odot}}
  -2\frac{\dot R_{\odot}}{R_{\odot}}
  + \frac{\dot T_{\odot}}{T_{\odot}} =
    \frac{1}{2} \frac{L_{\mbox{\scriptsize in}}}{\ell_{\odot}} 
\left(\frac{\dot M_{\odot}}{M_{\odot}} + \frac{\dot a}{a}\right)\,.
\end{equation}

Note that the sum of the angular momentum of inner planets amounts
only $0.16\%$ of the total $L$:
\begin{equation}
  \frac{L_{\mbox{\scriptsize in}}}{L} \simeq 1.6\times 10^{-3}. 
\end{equation}
Therefore, under the assumption that the change in the rotational
angular momentum of the Sun affects only the orbital angular
momentums of the inner planets, 
the required values which were calculated in the previous sections to
explain the secular increase of AU can now be revised to be $1.6\times
10^{-3}$
times smaller.  
In particular, the right hand side of Eq.~(\ref{eq:22}) is
\begin{equation}\label{eq:24}
  \frac{1}{2} \frac{L_{\mbox{\scriptsize in}}}{\ell_{\odot}}
  \frac{\dot a}{a}\simeq 
1.5\times 10^{-11}
\,(\mbox{cy}^{-1}). 
\end{equation}
Interestingly, it is the same order of magnitude as (actually it is
about 
1.6 times larger
 than) $\dot
  M_{\odot}/M_{\odot}$ (see Eq.~(\ref{eq:9.1})). 

Then we can conclude from Eq.~(\ref{eq:22}) that 
the decrease of rotational angular momentum of the Sun due to the
radiative mass loss has a 
significant contribution to the
secular increase of the orbital radius of the inner planets. 
%
\section{Conclusion}
In this paper, we considered the secular increase of astronomical 
unit recently reported by Krasinsky and Brumberg (2004), and 
suggested a possible explanation for this secular trend 
by means of the conservation law of total angular momentum. 
Assuming the existence of some tidal interactions that transfer the
angular momentum from the Sun to the planets system, we have obtained the
following results.

  From the reported value $\frac{d}{dt}\mbox{AU} = 15 \pm 4\,
\mbox{(m/cy)}$, we have obtained the required value for the variation of
rotational period of the Sun is about 
$21 \, \mbox{(ms/cy)}$,  if we assume
that eight planets in the solar system experience the same orbital
expansion rate.  
This value is sufficiently small, and at present it
seems there are no observational data which exclude this possibility.

Moreover, we have found that the effects of change in the moment of 
inertia of the Sun due to the radiative mass loss may be 
responsible for explaining the secular increase of AU. 
Especially, when we suppose that the orbital expansion occurs 
only in the inner planets region, 
the decrease of rotational angular momentum of the Sun has enough 
contribution to the secular increase of the orbital radius. 
Then as an answer to the question ``why is AU increasing ?'', 
we propose one possibility, namely 
``because the Sun is losing its angular momentum.''

As showed in previous section, the obtained value in
Eq.~(\ref{eq:24}) is practically about 1.6-times smaller than 
that due to $\dot M_{\odot}/M_{\odot}$. Nonetheless the estimated 
$d{\rm AU}/dt$ in the different data sets and fitting parameters 
is distributed within the range $7.9 \pm 0.2$ to 
$61.0 \pm 6.0~ {\rm (m/cy)}$ then seems not to be tightly 
constrained, see Table 2 of \citet{kb2004}. 
Hence we can say that the estimated value 
in Eq.~(\ref{eq:24}) falls into the suitable range. 

In the process of planetary ephemerides construction,
the effects due to both solar mass loss and rotational angular
momentum transfer are currently not included. However, 
as mentioned by \citet{noerdlinger2008}, 
the solar mass loss induces the variation in the orbital radius 
of planet $\delta \dot{r} \sim + 1~ {\rm (m/cy)}$. 
Furthermore as showed in this paper, the change of Sun's moment 
of inertia may also cause the significant contribution 
to orbital motion of planets. These facts may indicate the necessity
of reconsideration of the astronomical system of units since
AU is not only the conversion constant of two length units but also 
the value which characterizes $GM$ of the Sun via the 
following relation,
\begin{equation}\label{eq:def-aunits}
GM_{\rm Sun} = k^2 {\rm AU}^3/{\rm d}^2,
\end{equation}
where $k = 0.01720209895$ is the Gaussian gravitational constant
and ${\rm d}$ means a day so that ${\rm d} = 86400 ~({\rm s})$.
Until now, several authors discussed the redefinition of 
the system of astronomical units e.g., 
\citet{huang1995,standish2005,klioner2008,capitaine2009}.
By using Eqs. (\ref{eq:au1}) and (\ref{eq:def-aunits}), 
we obtain $GM_{\rm Sun}$ in SI units and so far $GM_{\rm Sun}$ 
is regarded conventionally as the ``fixed value'' in SI units.
However the discussions in \citet{noerdlinger2008} and present
paper imply that the effects of solar mass loss and the rotational 
angular momentum transfer of the Sun due to the radiative mass loss 
cannot be disregarded in terms of the growing observational accuracy 
and the definition of system of astronomical units. 

In present paper, we proposed the possible mechanism for explaining
the secular increase of AU, nonetheless we need to verify the 
validity of our model by means of the some tidal dissipation models 
of the Sun. 
Moreover
because the existence of $d{\rm AU}/dt$ is not confirmed robustly 
in terms of the independent analysis of observation by other 
ephemerides groups, it is important not only to perform the 
theoretical researches but also to re-analyze the data and to 
obtain more accurate value of $d{\rm AU}/dt$ adding new observations 
e.g., Mars Reconnaissance Orbiter, Phoenix and forthcoming MESSENGER 
which is cruising to the Mercury.
It also seems to be meaningful to use the observations of outer 
planets as well, such as Cassini, Pioneer 10/11, Voyager 1/2, and 
New Horizons for Pluto since it is more natural situation that 
the variation of moment of inertia of the Sun 
causes the orbital changes not only of inner
planets but also of outer ones.
Therefore in order to reveal the origin of secular increase of AU,
it is essential to investigate these subjects in detail.
\section*{Acknowledgments}
The authors are grateful to the referee for careful reading of
the manuscript and for fruitful comments and suggestions.
We also acknowledge to Prof. G. A. Krasinsky for
providing the information and comments about the AU issue.

\end{document}